\documentclass{aa} % for a normal version

\newcommand{\qvs}{2006~QV$_{89}$}
\newcommand{\qv}{\qvs\ }
\usepackage{graphicx}
\usepackage{hyperref}
\usepackage{xcolor}

\newcommand{\B}[1]{#1}

\newcommand{\okina}{`}
\newcommand{\Ou}{{\okina}Oumuamua}

\usepackage{txfonts}
\begin{document}

   \title{Elimination of a virtual impactor of 2006 QV$_{89}$ via deep non-detection}

   \author{      
    O.~R. Hainaut\inst{1}\fnmsep\thanks{\email{ohainaut@eso.org}}
    \and
    M. Micheli\inst{2}\fnmsep\thanks{\email{marco.micheli@esa.int}}
    \and
    J.~L. Cano\inst{2}
    \and
    J. Martín\inst{2}
    \and
    L. Faggioli\inst{2} % Laura
    \and
    R. Cennamo\inst{2} % Ramona
          }
   \institute{
             ESO, Karl-Schwarzschild-Stra\ss e 2, 85748 Garching-bei-M\"unchen, Germany\\
        \and
            ESA NEO Coordination Centre, Largo Galileo Galilei, 1, 00044 Frascati (RM), Italy\\
             }

   \date{Received 11 Jun. 2021; accepted 5 Aug. 2021}

% \abstract{}{}{}{}{} 
% 5 {} token are mandatory
 
  \abstract
   {As a consequence of the large (and growing) number of near-Earth objects discovered, some of them are lost before their orbit can be firmly established to ensure long-term recovery. A fraction of these present non-negligible chances of impact with the Earth. We present a method of targeted observations that allowed us to eliminate that risk by obtaining deep images of the area where the object would be, should it be on a collision orbit.
   }
   {\qv was one of these objects, with a chance of impact with the Earth on 2019 September~9. Its position uncertainty (of the order of $1\degr$) and faintness (below $V\sim 24$) made it a difficult candidate for a traditional direct recovery. However, the position of the virtual impactors could be determined with excellent accuracy.
   }
   {In July 2019, the virtual impactors of \qv were particularly well placed, with a very small uncertainty region, and an expected magnitude of $V < 26$. The area was imaged using ESO's Very Large Telescope, in the context of the ESA/ESO collaboration on Near-Earth Objects, resulting in very constraining a non-detection. 
   }
   {We eliminated the virtual impactor, even without effectively recovering \qvs, indicating that it did not represent a threat.   
   }
   {This method of deep non-detection of virtual impactors demonstrated a large potential to eliminate the threat of otherwise difficult to recover near-Earth objects.
   }


   \maketitle
%
%-------------------------------------------------------------------

\section{Introduction}

Every year, about 2000 new Near-Earth Objects (NEOs) are discovered, most of them by dedicated surveys\footnote{\url{https://minorplanetcenter.net//iau/lists/YearlyBreakdown.html}}. As soon as the discovery is confirmed and, usually, after a few days, the objects obtain formal provisional designations, and their orbits can be constrained. Together with the orbital calculations, it is possible to investigate if any of these objects also present a direct threat to our planet by finding possible orbits that are compatible with both the available observational data and correspond to an impact in the future.

On average, for about 96\% of the objects, all impact possibilities are quickly excluded. For the remaining 4\%, however, scenarios corresponding to a possible future impact remain. In these cases, the easiest way to clarify the situation is to perform additional follow-up observations and repeat the computations to \B{check if} future impacts are still compatible with the new data.

This iterative process should ideally continue until all possibilities of impacts are excluded or possibly confirmed. However, other factors come into play, especially related to the observability of the targets. NEOs are typically discovered when close to the Earth and, therefore, they often quickly recede from our planet, becoming fainter. To extend the observed arc as much as possible, larger telescopes can be used. This is the goal of the collaboration between the European Southern Observatory (ESO) and the European Space Agency (ESA) Planetary Defence Office \citep{Hainaut+14}, in which an 8.2 m unit from ESO's Very Large Telescope (VLT) is used to measure NEOs down to magnitude $\sim 27$ or beyond (e.g. 2021~GN$_2$ at $G\sim 27.7$ \footnote{MPEC 2021-K33: \url{https://www.minorplanetcenter.net/mpec/K21/K21K33.html}}). Eventually, however, the objects move out of reach of all telescopes. When that happens, no new data can be acquired, often for a very long time, until the object approaches the Earth again and can be recovered.

There are, however, some circumstances when no additional opportunities for observations may exist and, therefore, the possible threat cannot be excluded. The most likely cause is that an insufficient number of observations could be collected around and immediately after the time of discovery, and therefore our knowledge of the orbit of the object is not sufficient to meaningfully predict its position on the plane of the sky next time it becomes close enough to be observed. In this case, the object is effectively lost, and no additional observations can be obtained unless it is rediscovered by chance, and subsequently linked to the original detections.

For a threatening object, especially one with high probability impacts in the future, this situation is unacceptable. Fortunately, an additional approach to clarify the impact threat exists, even without effectively reobserving the lost object. The method, first proposed by \citet{2000Icar..145...12M}, is based on the idea that, to exclude a possible impact, it is only necessary to ensure that the asteroid is not located on the specific orbit (or family of similar orbits) that will lead it to collide with the Earth. It is usually possible to determine such trajectories very accurately since they are ``anchored'' by the strong constraint that they have to impact with the Earth at a specific future time. It is, therefore, possible to exclude (or possibly confirm) the impact by just observing the position in the sky where the object would appear if it were the impacting orbit. A solid non-detection of the object in that area would prove that the impact will not happen, even without detecting the object itself.

This technique has already been considered in the past, for example by \citet{2019Icar..317...39M}, who excluded a possible impact by analyzing existing archival data containing the location of a possible impacting trajectory. In this paper, we present a dedicated observation targeting a specific high-risk object. The method outlined here \B{forms the foundation for} a proposed protocol to design, execute and report such non-detections, in a rigorous way that can be properly taken into account by impact monitoring efforts.

%--------------------------------------------------------------------
\section{The target}\label{sec:target}

The target is NEO \qvs. Discovered on 2006 August~29 by the Catalina Sky Survey in Arizona, USA. Various facilities subsequently followed it up by until 2006 September~8. The limited observational coverage resulted in a poorly determined orbit. The orbit nevertheless revealed a close approach with the Earth, including a possible impact, on 2019 September~9.

\subsection{Orbit determination process}\label{sec:orbit_det}

The orbit determination process on which this analysis is based has been performed at ESA's NEO Coordination Centre, by using the AstOD software\footnote{AstOD was developed within the NEO Segment of ESA's Space Situational Awareness Programme.}. The resulting Keplerian orbital elements and their $1\sigma$ variation are shown in Table~\ref{tab:orbit}.
The dynamical model takes into account the $\mathcal{N}$-body gravitational attraction of major solar system objects (the Sun, eight planets, the Moon, the 16 most massive asteroids and Pluto), the parameterized post-Newtonian relativistic contributions and the oblateness of the Sun and the Earth. 
For the orbit determination process, the weighting scheme described in \citet{2017..296..139M} was adopted, and the outlier rejection procedure as described in \citet{Milani..book} was applied to the overall dataset. In the case of \qvs, the outlier rejection procedure discarded three observations, and the normalised astrometric Root Mean Square (RMS) residual after the differential corrections process was $\sim 0.58$.
The computed orbit resulting from the differential corrections, and its covariance matrix, formed the input for the skyprint analysis discussed in the next section.

%--------------------------------------------------- 
   \begin{table}
      \caption[]{Keplerian orbital elements of \qv at the epoch MJD $= 53978.0$}
         \label{tab:orbit}
         \begin{tabular}{cllc}
\hline
 Element    &Value          & $1\sigma$ variation  &Unit \\
 \hline
 $a$        &$1.1965$       &$1.78 \times 10^{-4}$  & au  \\
 $e$        &$0.2266$       &$1.44 \times 10^{-4}$  &      \\
 $i$        &$1.0690$       &$6.19 \times 10^{-4}$  & deg  \\
 $\Omega$   &$166.2189$     &$6.85 \times 10^{-4}$  & deg \\
 $\omega$   &$235.9946$     &$2.85 \times 10^{-3}$  & deg \\
 $M$        &$317.9547$     &$1.45 \times 10^{-2}$  & deg \\
\hline
         \end{tabular}
   \end{table}
%

%=========
\subsection{Position of the virtual impactors}\label{sec:position} 
%-------------------------
\begin{figure}
    \centering
    \includegraphics[width=8.5cm]{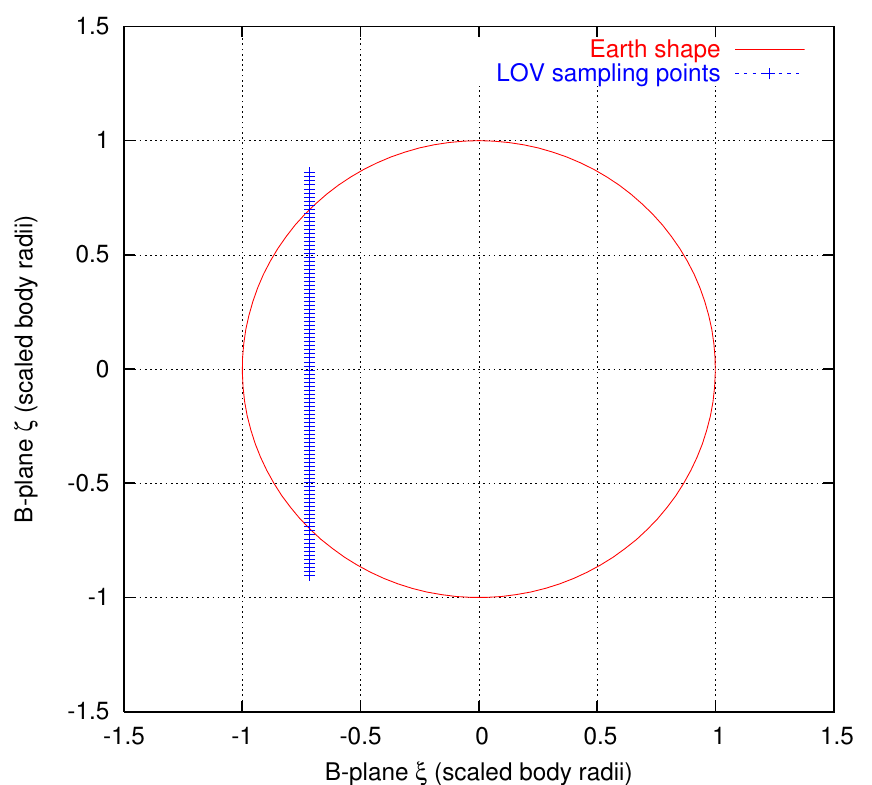}
    \caption{Intersection of the LoV with the cross-section of the Earth in the B-plane of the 2019 September encounter.}
    \label{fig:NIRAT}
\end{figure}
%-------------------------

%\B{JL: LOTNAV}%----------
Asteroid impact monitoring methods include several possible approaches, for example, the Line of Variations (LoV) approach \citep{2005Icar..173..362M} and Monte Carlo methods. 
%The impact risk of asteroid \qv was established soon after its discovery and was left mostly untouched as it remained unobserved since its discovery in 2006. 
The planned observations of \qv required the determination of the family of virtual asteroids that could impact the Earth, being at the same time compatible with the available observations and uncertainties. These are typically referred to as {\em Virtual Impactors} (VIs) and the set of those that showed an impact with Earth on 2019 September~9 were mapped to a given interval in the line of variations.

Starting from the computed orbit solution (Sect.~\ref{sec:orbit_det}), the mapping of the 2019 VIs in the LoV was performed in January of that year using the NIRAT software \citep{NIRAT}. The sampling of the LoV mapped to the impact plane \citep[as defined by ][]{2003A&A...408.1179V} in 2019 September is provided in Fig.~\ref{fig:NIRAT}. Those sampled solutions allowed us to determine the initial state vector of the asteroids that were at the impact limits with the Earth, and thus enabled at a later stage the computation of observational ephemerides any time before the impact chance in 2019 September.

Even if the second largest eigenvalue of the orbit determination covariance was very small compared to the largest one, we performed an additional two-dimensional elliptical sampling of the covariance. This allowed us to obtain a $3\sigma$ two-dimensional boundary on the same impact plane. However, negligible changes could be observed in the resulting limiting impact points, as expected by the large difference in the values of the two first eigenvalues. Consequently, we simply used the initial linear sampling of the LoV for all further computations.

%===========================
\subsection{Brightness of the virtual impactors}\label{sec:photometry} 
%\B{oli: photometry}

Based on the (notoriously inaccurate) photometric measurements reported at MPC, the absolute magnitude of \qv is $H=25.5$. \B{Modern, well-calibrated sky surveys have accurate photometry, so the quality of MPC magnitude is bound to improve.} \B{The slightly different geometry for the various orbits of the VIs (hence slightly different conversion from observed magnitude to $H$) does not have an impact on that value.}

Small NEOs are significantly elongated, resulting in large lightcurve amplitudes. \citet{Kwiatkowski+10} reported the amplitude of 12 such objects, with an average lightcurve range of 1.4~mag, the largest range being 2.2~mag. The extreme, probably pathological case of 1I/\Ou\ had a lightcurve range of 2.5~mag \citep{Meech+17}. The photometric measurements of NEOs provided with the astrometric reports are also likely to be biased towards the brighter part of the lightcurve, because they correspond to times when the object was easier, and therefore more likely, to be detected and measured. Overall, the object has therefore the potential to be significantly fainter than predicted from its published $H$. 

Correcting the absolute magnitude for the helio- and geocentric distances and for the solar phase effect (using a customary 0.03 mag/deg correction), we would expect  $V \sim 23.7$ at the time our observations were going to be scheduled. Accounting for the possible rotational variability and the uncertainty on $H$, we set the target magnitude to $V = 26$, i.e. over two magnitudes fainter than expected, ensuring the object be brighter than that limit. Coincidentally, this limit also corresponds to an object with a diameter of $\sim 10$ m, which would cause little or no harm in case of impact.

%===================================
\section{Observations} % Oli
The \qv VI was observed with Unit Telescope~1 of the ESO VLT on Paranal, Chile, using the FOcal Reducer and low dispersion Spectrograph 2 \citep[FORS2]{fors94}, without any filter in order to reach the deepest possible limiting magnitude. The instrument was equipped with its ``red'' CCD, a mosaic of two MIT/LL detectors. Only the main chip, ``A'', or CCID20-14-5-3, was used. Its effective field of view covers $7'\times4'$, with $0.252''$ binned pixels. 

The observations took place in Service Mode on 2019 July 4 and 5; the circumstances and geometry of the observations are listed in Table~\ref{tab:observations}. The data are publicly available from the ESO Science Archive, querying it\footnote{
\href{http://archive.eso.org/wdb/wdb/eso/eso_archive_main/query?wdbo=html\%2fdisplay&max_rows_returned=200&ob_id=2445558\%20|\%202445555}{Direct URL to the data at archive.eso.org}
}for Observation Blocks numbers (OB ID) 2445555 and 2445558.

%--------------------------------------------------- 
   \begin{table*}
      \caption[]{Geometry and circumstances of the observations of \qvs.}
         \label{tab:observations}

         \begin{tabular}{lccllllll}
\hline
 UT Date     &Start &End    & $r$  &$\Delta$& $\alpha$& N &t  &S \\
 \hline
 2019-07-04 &05:26 &06:37  & 1.252& 0.262  & 23.1    &45 &60 &0.87\\
 2019-07-05 &03:57 &05:11  & 1.249& 0.256  & 22.4    &44 &60 &0.91\\
\hline
         \end{tabular}
         
Notes: Start and End list the beginning and end UT of the observation sequence, $r$ and $\Delta$ the helio- and geocentric distances in au, $\alpha$ the solar phase [deg]. N is the number of exposures; t, the individual exposure time [s]; and S the seeing [arcsec].
   \end{table*}
   
%===============================================================
\section{Data processing}\label{sec:dataproc}
%-----------------------
\begin{figure*}
    \centering
    \includegraphics[width=18.5cm]{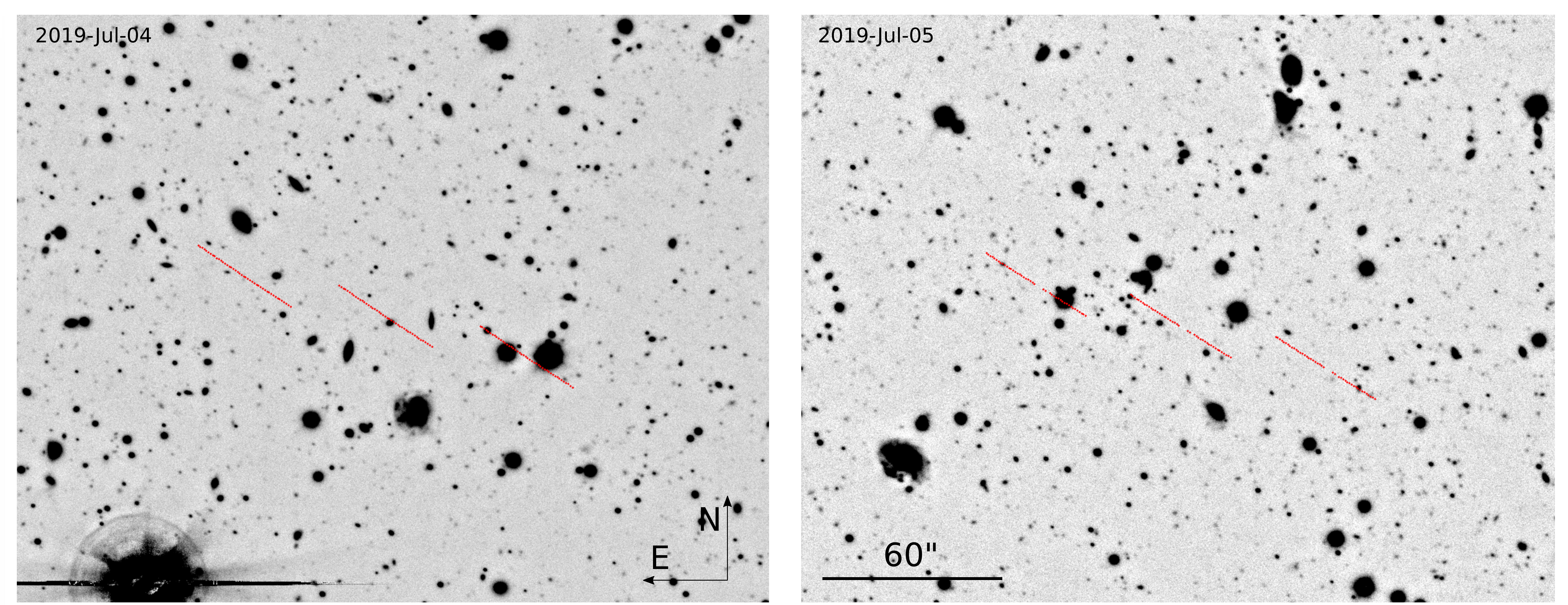}
    \caption{Stack of the images aligned on the stars. \B{The tracks for three VIs from the continuous swarm illustrated in Fig.~1 have been marked. They represent the most eastern and western VIs as well as the centre of the swarm. The eastern and western tracks bound the region to be searched}. The negative linear greyscale ranges from $2\sigma$ below to $15\sigma$ above the background.}
    \label{fig:stars}
\end{figure*}
%-----------------------
\begin{figure*}
    \centering
    \includegraphics[width=18.5cm]{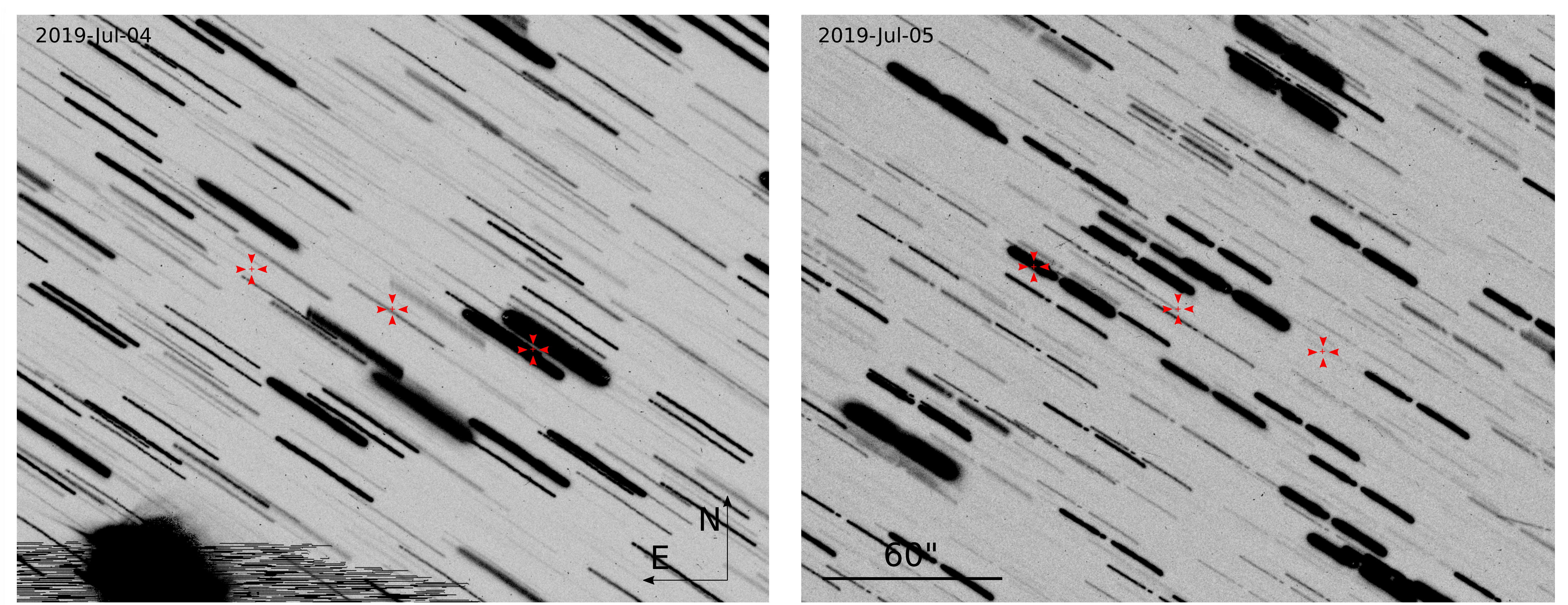}
    \caption{Stack of the images aligned on the expected position of the centre VI. \B{The position of the most eastern and most western VIs are also indicated. \qv appearing between these two extremes would be on a collision orbit.}
    The grey scale is the same as in Fig.~\ref{fig:stars}. }
    \label{fig:trails}
\end{figure*}
%-----------------------
\begin{figure*}
    \centering
    \includegraphics[width=18.5cm]{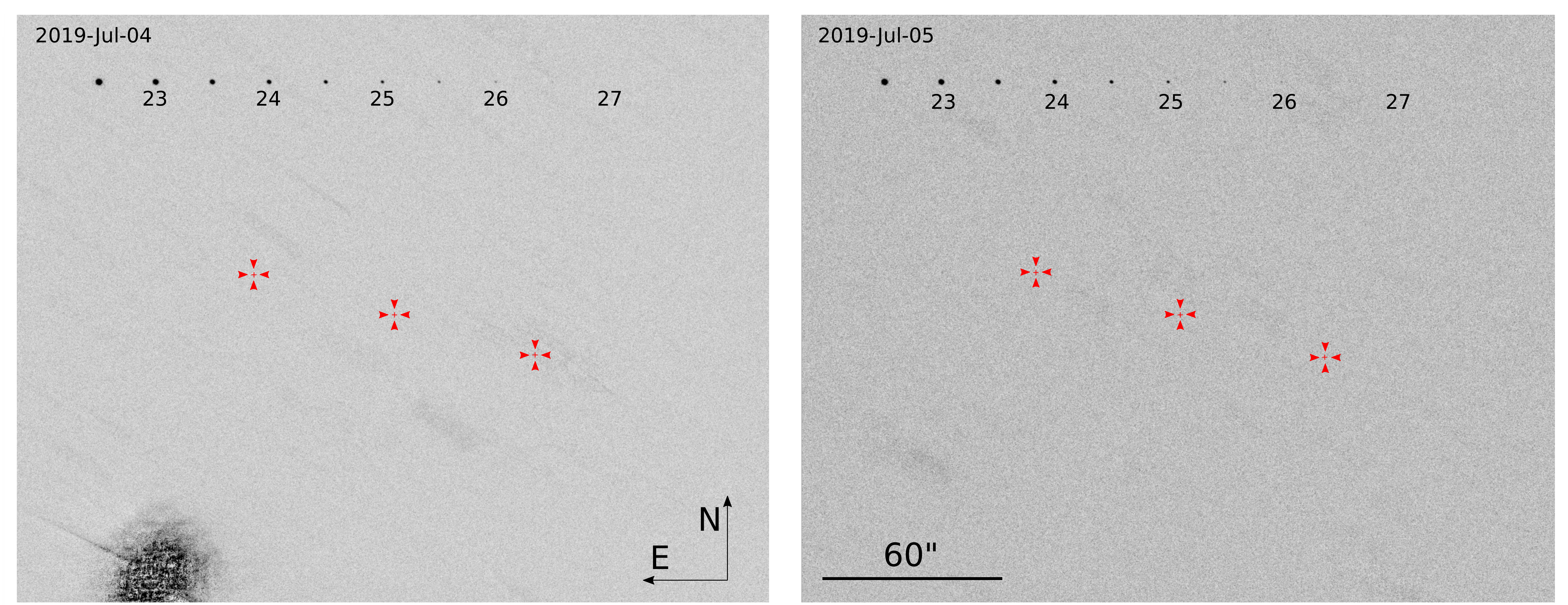}
    \caption{Stack of the background-subtracted frames, aligned on the expected position of the centre VI.  \B{The position of the most eastern and most western VIs are also indicated. \qv appearing between these two extremes would be on a collision orbit.} The grey scale is the same as in Fig.~\ref{fig:stars}. Artificial objects were introduced in the images; their magnitudes are indicated. As a reference, \qv is expected at mag $\sim 23.7$, and must be brighter than 26.    }
    \label{fig:klean}
\end{figure*}
%-----------------------

A bias template was obtained by averaging 14 zero-second exposures with outlier rejection; this template was subtracted from all frames. The level of the sky background on the \qv frames was high, about 7500 to 8800~adu. These images were normalised to a unit background by measuring the median sky level over the central region of the frame, where \qv was expected. The background objects were masked. The frames were averaged with outlier rejection, resulting in a template flat-field. As the observations were obtained offsetting the telescope between each exposure following a random pattern that ensured that each image was pointed at at least 2$''$ from any other image in the sequence, and because the field had only a low density of field stars, the resulting flat-field is extremely clean. Thanks to the 90 images participating in this template, the signal-to-noise ratio of the flat-field template was over 800. The original \qv frames were then divided by the flat-field template and normalised to 1s exposure time. The resulting images were sky-subtracted using SExtractor \citep{SEx}, then aligned on the first frame of the sequence using a dozen of field stars. For both nights, a background template was created by averaging the image from that night. These images, displayed in Fig.~\ref{fig:stars}, were used as astrometric reference. They were calibrated using over 20 stars from the Gaia DR2 catalogue \citep{Gaia16,Gaia18} using 2D second-degree polynomials. The RMS residual on the stars was better than $0.03''$. The expected pixel position of each of the \qv VIs in each image was computed using their ephemerides from Sect.~\ref{sec:position} and the astrometric calibration. Figure~\ref{fig:stars} shows these position in the context of the field objects, and Fig.~\ref{fig:trails} displays traditional stacks of the images aligned on the expected position of the object, with the stars trailed.

We used various methods to photometrically calibrate the frames.
First, the Gaia stars present in the frames were measured. From the few non-saturated stars, we estimated a zero point of 28.8. Second, we used the sky brightness. From the Paranal sky measurements in B, V, R, I \footnote{\url{https://www.eso.org/gen-fac/pubs/astclim/paranal/skybackground/}}, we estimated a sky brightness in filter-less observations of 19.7. From the sky brightness in individual frames, we measured a ZP=28.8. Alternatively, we measured the noise of the sky background. As that noise is dominated by the Poisson noise, we converted it into a brightness, which results in ZP=28.7. Third, we considered that the filter-less observations are similar to using the union of the B, V, R, and I filters, and combined the standard zero points obtained from Quality Control into ZP=28.5. The width of the B+V+R+I filters underestimate the width of the CCD passband and the transmission of the individual filters is less than 100\%, explaining why this ZP is lower than the other estimates. Overall, we keep ZP=28.7, with an uncertainty of $\sim 0.1$~mag.

We subtracted the background templates from the individual frames, resulting in images where the background objects were removed. Because of seeing variations and of the apparent rotation of diffraction patterns caused by the secondary mirror spider, this subtraction is far from perfect, in particular close to the core of each star. Nevertheless, the wider wings of the point spread function and the extended objects are very well subtracted. As the star templates are composed of 44 or 45 images, their noise level is over $6\times$ better than that of the individual images; this subtraction, therefore, does not significantly degrade the signal-to-noise ratio of the images. Also, as \qv moves from frame to frame, and because the star templates were obtained rejecting outliers, the (potential) signal from \qv is not affected. We co-added both the original frames and the background-subtracted frames (using either plain average, or an average with outlier rejection), resulting in the traditional image showing trailed stars, and a clean image that would show only the moving object. This method has been used, for instance, to detect comet 1P/Halley at magnitude 28.2 \citep{Hainaut+04}. The resulting images are shown in Fig.~\ref{fig:klean}. As a test, artificial objects covering a range of magnitudes were included in the frames; they appear as point sources at the top of the field.

We also produced additional image stacks using various subsets of the images (first and second half of the series, odd and even images). We carefully explored visually the region where the VI was expected, by blinking the various stacks. Faint candidates were investigated and found to be caused by star residuals. We could not identify any convincing candidate: the object is not detected.

%=====================================================================
\section{Quantifying the non-detection of \qv}
%--------------
\begin{figure*}
    \centering
    \includegraphics[width=18.5cm]{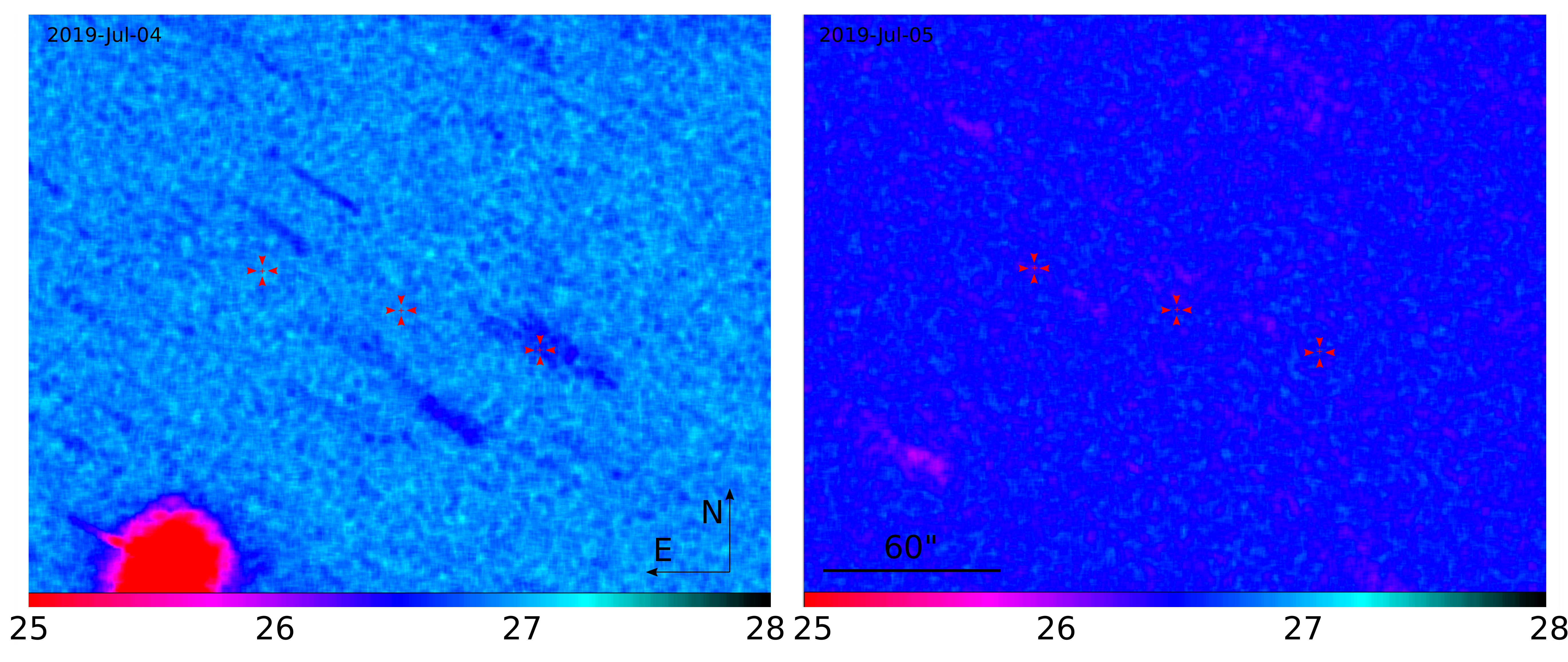}
    \caption{Map of the limiting magnitude of the stacks presented in Fig.~\ref{fig:klean}, at the $5\sigma$ level for a point source.  }
    \label{fig:lim}
\end{figure*}
%--------------
\begin{figure*}
    \centering
    \includegraphics[width=18.5cm]{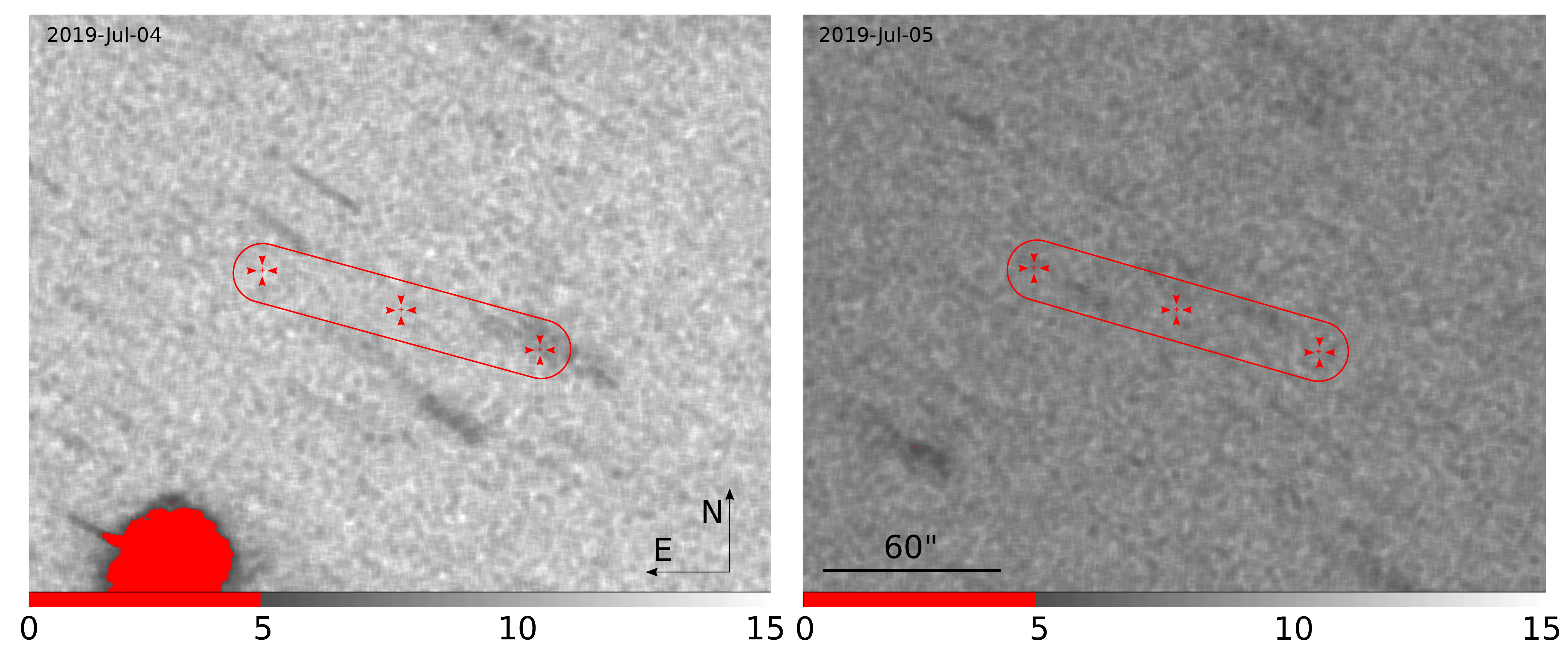}
    \caption{Map of the expected signal-to-noise ratio of the VI at its lowest magnitude 26. The regions where this SNR is below 5 are shaded in red. The zone of interest is marked by a red contour, $10''$ around the virtual impactors; \B{the positions of the central VI and the most eastern and western ones are marked}.  }
    \label{fig:sig}
\end{figure*}
%--------------

In order to quantify this non-detection, the maps of limiting magnitudes of the stacks were computed as follow. First, the residual background (at the 0.1 adu level) was subtracted using SExtractor, ensuring that the local mean background is zero.
The local noise of the image, 
\begin{equation}
    \sigma = \sqrt{\frac{ \sum{ (x_{ij} - \Bar{x})^2}}{N}}  ,
\end{equation}
was then estimated by filtering the squared background-subtracted image with a box 5~pixel wide (about $1''$). The value of a pixel is:
\begin{equation}
    f = \frac{ \sum{ x_{ij}^2 }}{N} =  \sigma^2  ,
\end{equation}
where $x_{ij}$ the flux of pixel $ij$ in the background-subtracted image,  and $N=25$ is the number of pixels in the filter box. By construction, the average value of $x_{ij}$ in the box is 0.

This $\sigma$ is then converted in the limiting magnitude for the $5\sigma$ detection of a point source in an $d=1'' = 3.97$~pixel aperture:
\begin{equation}
    M_{\rm lim} = {\rm ZP} -2.5 \log(5  d \sigma)~,
\end{equation}
where ZP is the photometric zero-point. 
Figure~\ref{fig:lim} shows the map of limiting magnitude. The brightest $5\sigma$ limits for point sources over the region of interest are 26.8 and 26.1 for each night respectively.

Similarly, a map of the significance level at which an object of magnitude $M$ would be measured is computed as 
\begin{equation}
    S = \frac{10^{-0.4(M-{\rm ZP})}}{d\sigma} ~.
\end{equation}

We used $ZP=28.7$ (from the previous section). Figure~\ref{fig:sig} shows the map of the expected signal-to-noise ratio for an object with $M=26$ (Sect.~\ref{sec:photometry}). The average expected signal-to-noise over the region of interest were 11.2 and 7.8 for both nights. Even considering the minimum value over the region of interest, 7.1 and 5.7, the object would have been brightly detected even at the very pessimistic brightness.

%==================================================
\section{Conclusions}
   \begin{enumerate}
      \item The orbit of NEO \qv was poorly known, but showed chances of collision with the Earth on 2019 September~9. The ephemerides of the virtual impactors were computed. 
      \item The magnitude of \qv expected from the published photometry is $V=23.7$. Accounting for the uncertainties on these measurements, and for a possible shape-induced variability, we set $V<26$ as the realistic absolute limit beyond which \qv cannot realistically be. 
      \item We obtained deep observations on two nights with FORS2 on the VLT, on the field where the \qv virtual impactors were expected, reaching limiting magnitude 26.8 and 26.5 ($5\sigma$ for point sources) on average over the area of interest.
      \item The individual images and stack of frames were carefully examined, resulting in constraining non-detection of the object. At a magnitude of 26, the object would have been detected with a signal-to-noise ratio (SNR) of 11.2 and 7.8 (for each night) on average over the region of interest. The shallowest detections would have been with SNR of 7.1 and 5.7.
      \item We can therefore rule out the presence of \qv in the VI area; the probability for it to be present but not detected is below $(1 - P(7.1\sigma)) \sim 0.000\,000\,000\,256\%$ in the most pessimistic case where its magnitude would be 26, i.e. over two magnitudes fainter than expected. As a consequence, even though \qv was not recovered, we could be sure that it would not hit the Earth on 2019 September~9.
   \end{enumerate}

Given the importance of \qv and the attention attracted to the target by this campaign and the possible detectability of the object in 2019 July, a dedicated effort to attempt its actual recovery was carried out by David J. Tholen, using the wide-field MegaCam imager on the Canada-France-Hawaii Telescope (CFHT). Thanks to the wide field of MegaCam, he successfully recovered the object on 2019 July 14, finding it about 3 degrees away from the nominal position, \B{and within 0.1 mag from the expected brightness}. His recovery was subsequently confirmed by our team using the Schmidt telescope at Calar Alto, Spain, and published by the Minor Planet Center\footnote{MPEC 2019-P85: \url{https://www.minorplanetcenter.net/mpec/K19/K19P85.html}}.

The observational procedure presented in this work outlines a rigorous process to apply the concepts of \citet{2005Icar..173..362M} to an actual threatening object. The main goal of this work is to be used as a reference case for the development of a formal and internationally agreed protocol to eradicate threatening VIs of lost objects via negative observations. This protocol is being developed at the time of preparation of this paper (early 2021). Once completed, it will form the basis on which negative observations can be formally recognised and integrated into the impact monitoring activities carried out by the various international dedicated centres. 

\begin{acknowledgements}
This work is based on observations collected at the European Southern Observatory under ESO programme 5101.C-0575. 
The authors are grateful to the Paranal staff for their dedication in obtaining our NEO observations for this programme.
This work has made use of data from the European Space Agency (ESA) mission
{\it Gaia} (\href{https://www.cosmos.esa.int/gaia}), processed by the {\it Gaia}
Data Processing and Analysis Consortium (DPAC,
\url{https://www.cosmos.esa.int/web/gaia/dpac/consortium}). Funding for the DPAC
has been provided by national institutions, in particular the institutions
participating in the {\it Gaia} Multilateral Agreement.
      
\end{acknowledgements}


\bibliographystyle{aa}
\bibliography{biblio}

\begin{thebibliography}{15}
\expandafter\ifx\csname natexlab\endcsname\relax\def\natexlab#1{#1}\fi

\bibitem[{{Appenzeller} {et~al.}(1998){Appenzeller}, {Fricke}, {F{\"u}rtig},
  {G{\"a}ssler}, {H{\"a}fner}, {Harke}, {Hess}, {Hummel}, {J{\"u}rgens},
  {Kudritzki}, {Mantel}, {Meisl}, {Muschielok}, {Nicklas}, {Rupprecht},
  {Seifert}, {Stahl}, {Szeifert}, \& {Tarantik}}]{fors94}
{Appenzeller}, I., {Fricke}, K., {F{\"u}rtig}, W., {et~al.} 1998, The
  Messenger, 94, 1

\bibitem[{{Bertin} \& {Arnouts}(1996)}]{SEx}
{Bertin}, E. \& {Arnouts}, S. 1996, Astronomy and Astrophysics Supplement
  Series, 117, 393

\bibitem[{{Cano} {et~al.}(2013){Cano}, {Bellei}, \& {Mart\'{i}n}}]{NIRAT}
{Cano}, J.~L., {Bellei}, G., \& {Mart\'{i}n}, J. 2013, in International
  Astronautical Federation, Vol.~6

\bibitem[{{Gaia Collaboration} {et~al.}(2018){Gaia Collaboration}, {Brown},
  {Vallenari}, {Prusti}, {de Bruijne}, {Babusiaux}, {Bailer-Jones}, {Biermann},
  {Evans}, {Eyer}, {Jansen}, {Jordi}, {Klioner}, {Lammers}, {Lindegren},
  {Luri}, {Mignard}, {Panem}, {Pourbaix}, {Randich}, {Sartoretti}, {Siddiqui},
  {Soubiran}, {van Leeuwen}, {Walton}, {Arenou}, {Bastian}, {Cropper},
  {Drimmel}, {Katz}, {Lattanzi}, {Bakker}, {Cacciari}, {Casta{\~n}eda},
  {Chaoul}, {Cheek}, {De Angeli}, {Fabricius}, {Guerra}, {Holl}, {Masana},
  {Messineo}, {Mowlavi}, {Nienartowicz}, {Panuzzo}, {Portell}, {Riello},
  {Seabroke}, {Tanga}, {Th{\'e}venin}, {Gracia-Abril}, {Comoretto},
  {Garcia-Reinaldos}, {Teyssier}, {Altmann}, {Andrae}, {Audard},
  {Bellas-Velidis}, {Benson}, {Berthier}, {Blomme}, {Burgess}, {Busso},
  {Carry}, {Cellino}, {Clementini}, {Clotet}, {Creevey}, {Davidson}, {De
  Ridder}, {Delchambre}, {Dell'Oro}, {Ducourant},
  {Fern{\'a}ndez-Hern{\'a}ndez}, {Fouesneau}, {Fr{\'e}mat}, {Galluccio},
  {Garc{\'\i}a-Torres}, {Gonz{\'a}lez-N{\'u}{\~n}ez}, {Gonz{\'a}lez-Vidal},
  {Gosset}, {Guy}, {Halbwachs}, {Hambly}, {Harrison}, {Hern{\'a}ndez},
  {Hestroffer}, {Hodgkin}, {Hutton}, {Jasniewicz}, {Jean-Antoine-Piccolo},
  {Jordan}, {Korn}, {Krone-Martins}, {Lanzafame}, {Lebzelter}, {L{\"o}ffler},
  {Manteiga}, {Marrese}, {Mart{\'\i}n-Fleitas}, {Moitinho}, {Mora}, {Muinonen},
  {Osinde}, {Pancino}, {Pauwels}, {Petit}, {Recio-Blanco}, {Richards},
  {Rimoldini}, {Robin}, {Sarro}, {Siopis}, {Smith}, {Sozzetti}, {S{\"u}veges},
  {Torra}, {van Reeven}, {Abbas}, {Abreu Aramburu}, {Accart}, {Aerts},
  {Altavilla}, {{\'A}lvarez}, {Alvarez}, {Alves}, {Anderson}, {Andrei},
  {Anglada Varela}, {Antiche}, {Antoja}, {Arcay}, {Astraatmadja}, {Bach},
  {Baker}, {Balaguer-N{\'u}{\~n}ez}, {Balm}, {Barache}, {Barata}, {Barbato},
  {Barblan}, {Barklem}, {Barrado}, {Barros}, {Barstow}, {Bartholom{\'e}
  Mu{\~n}oz}, {Bassilana}, {Becciani}, {Bellazzini}, {Berihuete}, {Bertone},
  {Bianchi}, {Bienaym{\'e}}, {Blanco-Cuaresma}, {Boch}, {Boeche}, {Bombrun},
  {Borrachero}, {Bossini}, {Bouquillon}, {Bourda}, {Bragaglia}, {Bramante},
  {Breddels}, {Bressan}, {Brouillet}, {Br{\"u}semeister}, {Brugaletta},
  {Bucciarelli}, {Burlacu}, {Busonero}, {Butkevich}, {Buzzi}, {Caffau},
  {Cancelliere}, {Cannizzaro}, {Cantat-Gaudin}, {Carballo}, {Carlucci},
  {Carrasco}, {Casamiquela}, {Castellani}, {Castro-Ginard}, {Charlot},
  {Chemin}, {Chiavassa}, {Cocozza}, {Costigan}, {Cowell}, {Crifo}, {Crosta},
  {Crowley}, {Cuypers}, {Dafonte}, {Damerdji}, {Dapergolas}, {David}, {David},
  {de Laverny}, {De Luise}, {De March}, {de Martino}, {de Souza}, {de Torres},
  {Debosscher}, {del Pozo}, {Delbo}, {Delgado}, {Delgado}, {Di Matteo},
  {Diakite}, {Diener}, {Distefano}, {Dolding}, {Drazinos}, {Dur{\'a}n},
  {Edvardsson}, {Enke}, {Eriksson}, {Esquej}, {Eynard Bontemps}, {Fabre},
  {Fabrizio}, {Faigler}, {Falc{\~a}o}, {Farr{\`a}s Casas}, {Federici},
  {Fedorets}, {Fernique}, {Figueras}, {Filippi}, {Findeisen}, {Fonti},
  {Fraile}, {Fraser}, {Fr{\'e}zouls}, {Gai}, {Galleti}, {Garabato},
  {Garc{\'\i}a-Sedano}, {Garofalo}, {Garralda}, {Gavel}, {Gavras}, {Gerssen},
  {Geyer}, {Giacobbe}, {Gilmore}, {Girona}, {Giuffrida}, {Glass}, {Gomes},
  {Granvik}, {Gueguen}, {Guerrier}, {Guiraud}, {Guti{\'e}rrez-S{\'a}nchez},
  {Haigron}, {Hatzidimitriou}, {Hauser}, {Haywood}, {Heiter}, {Helmi}, {Heu},
  {Hilger}, {Hobbs}, {Hofmann}, {Holland}, {Huckle}, {Hypki}, {Icardi},
  {Jan{\ss}en}, {Jevardat de Fombelle}, {Jonker}, {Juh{\'a}sz}, {Julbe},
  {Karampelas}, {Kewley}, {Klar}, {Kochoska}, {Kohley}, {Kolenberg},
  {Kontizas}, {Kontizas}, {Koposov}, {Kordopatis}, {Kostrzewa-Rutkowska},
  {Koubsky}, {Lambert}, {Lanza}, {Lasne}, {Lavigne}, {Le Fustec}, {Le
  Poncin-Lafitte}, {Lebreton}, {Leccia}, {Leclerc}, {Lecoeur-Taibi},
  {Lenhardt}, {Leroux}, {Liao}, {Licata}, {Lindstr{\o}m}, {Lister}, {Livanou},
  {Lobel}, {L{\'o}pez}, {Managau}, {Mann}, {Mantelet}, {Marchal}, {Marchant},
  {Marconi}, {Marinoni}, {Marschalk{\'o}}, {Marshall}, {Martino}, {Marton},
  {Mary}, {Massari}, {Matijevi{\v{c}}}, {Mazeh}, {McMillan}, {Messina},
  {Michalik}, {Millar}, {Molina}, {Molinaro}, {Moln{\'a}r}, {Montegriffo},
  {Mor}, {Morbidelli}, {Morel}, {Morris}, {Mulone}, {Muraveva}, {Musella},
  {Nelemans}, {Nicastro}, {Noval}, {O'Mullane}, {Ord{\'e}novic},
  {Ord{\'o}{\~n}ez-Blanco}, {Osborne}, {Pagani}, {Pagano}, {Pailler},
  {Palacin}, {Palaversa}, {Panahi}, {Pawlak}, {Piersimoni}, {Pineau}, {Plachy},
  {Plum}, {Poggio}, {Poujoulet}, {Pr{\v{s}}a}, {Pulone}, {Racero}, {Ragaini},
  {Rambaux}, {Ramos-Lerate}, {Regibo}, {Reyl{\'e}}, {Riclet}, {Ripepi}, {Riva},
  {Rivard}, {Rixon}, {Roegiers}, {Roelens}, {Romero-G{\'o}mez}, {Rowell},
  {Royer}, {Ruiz-Dern}, {Sadowski}, {Sagrist{\`a} Sell{\'e}s}, {Sahlmann},
  {Salgado}, {Salguero}, {Sanna}, {Santana-Ros}, {Sarasso}, {Savietto},
  {Schultheis}, {Sciacca}, {Segol}, {Segovia}, {S{\'e}gransan}, {Shih},
  {Siltala}, {Silva}, {Smart}, {Smith}, {Solano}, {Solitro}, {Sordo}, {Soria
  Nieto}, {Souchay}, {Spagna}, {Spoto}, {Stampa}, {Steele},
  {Steidelm{\"u}ller}, {Stephenson}, {Stoev}, {Suess}, {Surdej}, {Szabados},
  {Szegedi-Elek}, {Tapiador}, {Taris}, {Tauran}, {Taylor}, {Teixeira},
  {Terrett}, {Teyssandier}, {Thuillot}, {Titarenko}, {Torra Clotet}, {Turon},
  {Ulla}, {Utrilla}, {Uzzi}, {Vaillant}, {Valentini}, {Valette}, {van Elteren},
  {Van Hemelryck}, {van Leeuwen}, {Vaschetto}, {Vecchiato}, {Veljanoski},
  {Viala}, {Vicente}, {Vogt}, {von Essen}, {Voss}, {Votruba}, {Voutsinas},
  {Walmsley}, {Weiler}, {Wertz}, {Wevers}, {Wyrzykowski}, {Yoldas},
  {{\v{Z}}erjal}, {Ziaeepour}, {Zorec}, {Zschocke}, {Zucker}, {Zurbach}, \&
  {Zwitter}}]{Gaia18}
{Gaia Collaboration}, {Brown}, A.~G.~A., {Vallenari}, A., {et~al.} 2018, \aap,
  616, A1

\bibitem[{{Gaia Collaboration} {et~al.}(2016){Gaia Collaboration}, {Prusti},
  {de Bruijne}, {Brown}, {Vallenari}, {Babusiaux}, {Bailer-Jones}, {Bastian},
  {Biermann}, {Evans}, {Eyer}, {Jansen}, {Jordi}, {Klioner}, {Lammers},
  {Lindegren}, {Luri}, {Mignard}, {Milligan}, {Panem}, {Poinsignon},
  {Pourbaix}, {Randich}, {Sarri}, {Sartoretti}, {Siddiqui}, {Soubiran},
  {Valette}, {van Leeuwen}, {Walton}, {Aerts}, {Arenou}, {Cropper}, {Drimmel},
  {H{\o}g}, {Katz}, {Lattanzi}, {O'Mullane}, {Grebel}, {Holland}, {Huc},
  {Passot}, {Bramante}, {Cacciari}, {Casta{\~n}eda}, {Chaoul}, {Cheek}, {De
  Angeli}, {Fabricius}, {Guerra}, {Hern{\'a}ndez}, {Jean-Antoine-Piccolo},
  {Masana}, {Messineo}, {Mowlavi}, {Nienartowicz}, {Ord{\'o}{\~n}ez-Blanco},
  {Panuzzo}, {Portell}, {Richards}, {Riello}, {Seabroke}, {Tanga},
  {Th{\'e}venin}, {Torra}, {Els}, {Gracia-Abril}, {Comoretto},
  {Garcia-Reinaldos}, {Lock}, {Mercier}, {Altmann}, {Andrae}, {Astraatmadja},
  {Bellas-Velidis}, {Benson}, {Berthier}, {Blomme}, {Busso}, {Carry},
  {Cellino}, {Clementini}, {Cowell}, {Creevey}, {Cuypers}, {Davidson}, {De
  Ridder}, {de Torres}, {Delchambre}, {Dell'Oro}, {Ducourant}, {Fr{\'e}mat},
  {Garc{\'\i}a-Torres}, {Gosset}, {Halbwachs}, {Hambly}, {Harrison}, {Hauser},
  {Hestroffer}, {Hodgkin}, {Huckle}, {Hutton}, {Jasniewicz}, {Jordan},
  {Kontizas}, {Korn}, {Lanzafame}, {Manteiga}, {Moitinho}, {Muinonen},
  {Osinde}, {Pancino}, {Pauwels}, {Petit}, {Recio-Blanco}, {Robin}, {Sarro},
  {Siopis}, {Smith}, {Smith}, {Sozzetti}, {Thuillot}, {van Reeven}, {Viala},
  {Abbas}, {Abreu Aramburu}, {Accart}, {Aguado}, {Allan}, {Allasia},
  {Altavilla}, {{\'A}lvarez}, {Alves}, {Anderson}, {Andrei}, {Anglada Varela},
  {Antiche}, {Antoja}, {Ant{\'o}n}, {Arcay}, {Atzei}, {Ayache}, {Bach},
  {Baker}, {Balaguer-N{\'u}{\~n}ez}, {Barache}, {Barata}, {Barbier}, {Barblan},
  {Baroni}, {Barrado y Navascu{\'e}s}, {Barros}, {Barstow}, {Becciani},
  {Bellazzini}, {Bellei}, {Bello Garc{\'\i}a}, {Belokurov}, {Bendjoya},
  {Berihuete}, {Bianchi}, {Bienaym{\'e}}, {Billebaud}, {Blagorodnova},
  {Blanco-Cuaresma}, {Boch}, {Bombrun}, {Borrachero}, {Bouquillon}, {Bourda},
  {Bouy}, {Bragaglia}, {Breddels}, {Brouillet}, {Br{\"u}semeister},
  {Bucciarelli}, {Budnik}, {Burgess}, {Burgon}, {Burlacu}, {Busonero}, {Buzzi},
  {Caffau}, {Cambras}, {Campbell}, {Cancelliere}, {Cantat-Gaudin}, {Carlucci},
  {Carrasco}, {Castellani}, {Charlot}, {Charnas}, {Charvet}, {Chassat},
  {Chiavassa}, {Clotet}, {Cocozza}, {Collins}, {Collins}, {Costigan}, {Crifo},
  {Cross}, {Crosta}, {Crowley}, {Dafonte}, {Damerdji}, {Dapergolas}, {David},
  {David}, {De Cat}, {de Felice}, {de Laverny}, {De Luise}, {De March}, {de
  Martino}, {de Souza}, {Debosscher}, {del Pozo}, {Delbo}, {Delgado},
  {Delgado}, {di Marco}, {Di Matteo}, {Diakite}, {Distefano}, {Dolding}, {Dos
  Anjos}, {Drazinos}, {Dur{\'a}n}, {Dzigan}, {Ecale}, {Edvardsson}, {Enke},
  {Erdmann}, {Escolar}, {Espina}, {Evans}, {Eynard Bontemps}, {Fabre},
  {Fabrizio}, {Faigler}, {Falc{\~a}o}, {Farr{\`a}s Casas}, {Faye}, {Federici},
  {Fedorets}, {Fern{\'a}ndez-Hern{\'a}ndez}, {Fernique}, {Fienga}, {Figueras},
  {Filippi}, {Findeisen}, {Fonti}, {Fouesneau}, {Fraile}, {Fraser}, {Fuchs},
  {Furnell}, {Gai}, {Galleti}, {Galluccio}, {Garabato}, {Garc{\'\i}a-Sedano},
  {Gar{\'e}}, {Garofalo}, {Garralda}, {Gavras}, {Gerssen}, {Geyer}, {Gilmore},
  {Girona}, {Giuffrida}, {Gomes}, {Gonz{\'a}lez-Marcos},
  {Gonz{\'a}lez-N{\'u}{\~n}ez}, {Gonz{\'a}lez-Vidal}, {Granvik}, {Guerrier},
  {Guillout}, {Guiraud}, {G{\'u}rpide}, {Guti{\'e}rrez-S{\'a}nchez}, {Guy},
  {Haigron}, {Hatzidimitriou}, {Haywood}, {Heiter}, {Helmi}, {Hobbs},
  {Hofmann}, {Holl}, {Holland}, {Hunt}, {Hypki}, {Icardi}, {Irwin}, {Jevardat
  de Fombelle}, {Jofr{\'e}}, {Jonker}, {Jorissen}, {Julbe}, {Karampelas},
  {Kochoska}, {Kohley}, {Kolenberg}, {Kontizas}, {Koposov}, {Kordopatis},
  {Koubsky}, {Kowalczyk}, {Krone-Martins}, {Kudryashova}, {Kull}, {Bachchan},
  {Lacoste-Seris}, {Lanza}, {Lavigne}, {Le Poncin-Lafitte}, {Lebreton},
  {Lebzelter}, {Leccia}, {Leclerc}, {Lecoeur-Taibi}, {Lemaitre}, {Lenhardt},
  {Leroux}, {Liao}, {Licata}, {Lindstr{\o}m}, {Lister}, {Livanou}, {Lobel},
  {L{\"o}ffler}, {L{\'o}pez}, {Lopez-Lozano}, {Lorenz}, {Loureiro},
  {MacDonald}, {Magalh{\~a}es Fernandes}, {Managau}, {Mann}, {Mantelet},
  {Marchal}, {Marchant}, {Marconi}, {Marie}, {Marinoni}, {Marrese},
  {Marschalk{\'o}}, {Marshall}, {Mart{\'\i}n-Fleitas}, {Martino}, {Mary},
  {Matijevi{\v{c}}}, {Mazeh}, {McMillan}, {Messina}, {Mestre}, {Michalik},
  {Millar}, {Miranda}, {Molina}, {Molinaro}, {Molinaro}, {Moln{\'a}r},
  {Moniez}, {Montegriffo}, {Monteiro}, {Mor}, {Mora}, {Morbidelli}, {Morel},
  {Morgenthaler}, {Morley}, {Morris}, {Mulone}, {Muraveva}, {Musella},
  {Narbonne}, {Nelemans}, {Nicastro}, {Noval}, {Ord{\'e}novic},
  {Ordieres-Mer{\'e}}, {Osborne}, {Pagani}, {Pagano}, {Pailler}, {Palacin},
  {Palaversa}, {Parsons}, {Paulsen}, {Pecoraro}, {Pedrosa}, {Pentik{\"a}inen},
  {Pereira}, {Pichon}, {Piersimoni}, {Pineau}, {Plachy}, {Plum}, {Poujoulet},
  {Pr{\v{s}}a}, {Pulone}, {Ragaini}, {Rago}, {Rambaux}, {Ramos-Lerate},
  {Ranalli}, {Rauw}, {Read}, {Regibo}, {Renk}, {Reyl{\'e}}, {Ribeiro},
  {Rimoldini}, {Ripepi}, {Riva}, {Rixon}, {Roelens}, {Romero-G{\'o}mez},
  {Rowell}, {Royer}, {Rudolph}, {Ruiz-Dern}, {Sadowski}, {Sagrist{\`a}
  Sell{\'e}s}, {Sahlmann}, {Salgado}, {Salguero}, {Sarasso}, {Savietto},
  {Schnorhk}, {Schultheis}, {Sciacca}, {Segol}, {Segovia}, {Segransan},
  {Serpell}, {Shih}, {Smareglia}, {Smart}, {Smith}, {Solano}, {Solitro},
  {Sordo}, {Soria Nieto}, {Souchay}, {Spagna}, {Spoto}, {Stampa}, {Steele},
  {Steidelm{\"u}ller}, {Stephenson}, {Stoev}, {Suess}, {S{\"u}veges}, {Surdej},
  {Szabados}, {Szegedi-Elek}, {Tapiador}, {Taris}, {Tauran}, {Taylor},
  {Teixeira}, {Terrett}, {Tingley}, {Trager}, {Turon}, {Ulla}, {Utrilla},
  {Valentini}, {van Elteren}, {Van Hemelryck}, {van Leeuwen}, {Varadi},
  {Vecchiato}, {Veljanoski}, {Via}, {Vicente}, {Vogt}, {Voss}, {Votruba},
  {Voutsinas}, {Walmsley}, {Weiler}, {Weingrill}, {Werner}, {Wevers},
  {Whitehead}, {Wyrzykowski}, {Yoldas}, {{\v{Z}}erjal}, {Zucker}, {Zurbach},
  {Zwitter}, {Alecu}, {Allen}, {Allende Prieto}, {Amorim},
  {Anglada-Escud{\'e}}, {Arsenijevic}, {Azaz}, {Balm}, {Beck}, {Bernstein},
  {Bigot}, {Bijaoui}, {Blasco}, {Bonfigli}, {Bono}, {Boudreault}, {Bressan},
  {Brown}, {Brunet}, {Bunclark}, {Buonanno}, {Butkevich}, {Carret}, {Carrion},
  {Chemin}, {Ch{\'e}reau}, {Corcione}, {Darmigny}, {de Boer}, {de Teodoro}, {de
  Zeeuw}, {Delle Luche}, {Domingues}, {Dubath}, {Fodor}, {Fr{\'e}zouls},
  {Fries}, {Fustes}, {Fyfe}, {Gallardo}, {Gallegos}, {Gardiol}, {Gebran},
  {Gomboc}, {G{\'o}mez}, {Grux}, {Gueguen}, {Heyrovsky}, {Hoar}, {Iannicola},
  {Isasi Parache}, {Janotto}, {Joliet}, {Jonckheere}, {Keil}, {Kim},
  {Klagyivik}, {Klar}, {Knude}, {Kochukhov}, {Kolka}, {Kos}, {Kutka}, {Lainey},
  {LeBouquin}, {Liu}, {Loreggia}, {Makarov}, {Marseille}, {Martayan},
  {Martinez-Rubi}, {Massart}, {Meynadier}, {Mignot}, {Munari}, {Nguyen},
  {Nordlander}, {Ocvirk}, {O'Flaherty}, {Olias Sanz}, {Ortiz}, {Osorio},
  {Oszkiewicz}, {Ouzounis}, {Palmer}, {Park}, {Pasquato}, {Peltzer}, {Peralta},
  {P{\'e}turaud}, {Pieniluoma}, {Pigozzi}, {Poels}, {Prat}, {Prod'homme},
  {Raison}, {Rebordao}, {Risquez}, {Rocca-Volmerange}, {Rosen}, {Ruiz-Fuertes},
  {Russo}, {Sembay}, {Serraller Vizcaino}, {Short}, {Siebert}, {Silva},
  {Sinachopoulos}, {Slezak}, {Soffel}, {Sosnowska}, {Strai{\v{z}}ys}, {ter
  Linden}, {Terrell}, {Theil}, {Tiede}, {Troisi}, {Tsalmantza}, {Tur},
  {Vaccari}, {Vachier}, {Valles}, {Van Hamme}, {Veltz}, {Virtanen}, {Wallut},
  {Wichmann}, {Wilkinson}, {Ziaeepour}, \& {Zschocke}}]{Gaia16}
{Gaia Collaboration}, {Prusti}, T., {de Bruijne}, J.~H.~J., {et~al.} 2016,
  \aap, 595, A1

\bibitem[{{Hainaut} {et~al.}(2014){Hainaut}, {Koschny}, \&
  {Micheli}}]{Hainaut+14}
{Hainaut}, O., {Koschny}, D., \& {Micheli}, M. 2014, in Asteroids, Comets,
  Meteors 2014, ed. K.~{Muinonen}, A.~{Penttil{\"a}}, M.~{Granvik},
  A.~{Virkki}, G.~{Fedorets}, O.~{Wilkman}, \& T.~{Kohout}, 197

\bibitem[{{Hainaut} {et~al.}(2004){Hainaut}, {Delsanti}, {Meech}, \&
  {West}}]{Hainaut+04}
{Hainaut}, O.~R., {Delsanti}, A., {Meech}, K.~J., \& {West}, R.~M. 2004, \aap,
  417, 1159

\bibitem[{{Kwiatkowski} {et~al.}(2010){Kwiatkowski}, {Polinska}, {Loaring},
  {Buckley}, {O'Donoghue}, {Kniazev}, \& {Romero Colmenero}}]{Kwiatkowski+10}
{Kwiatkowski}, T., {Polinska}, M., {Loaring}, N., {et~al.} 2010, \aap, 511, A49

\bibitem[{{Meech} {et~al.}(2017){Meech}, {Weryk}, {Micheli}, {Kleyna},
  {Hainaut}, {Jedicke}, {Wainscoat}, {Chambers}, {Keane}, {Petric}, {Denneau},
  {Magnier}, {Berger}, {Huber}, {Flewelling}, {Waters}, {Schunova-Lilly}, \&
  {Chastel}}]{Meech+17}
{Meech}, K.~J., {Weryk}, R., {Micheli}, M., {et~al.} 2017, \nat, 552, 378

\bibitem[{{Micheli} {et~al.}(2019){Micheli}, {Cano}, {Faggioli}, {Ceccaroni},
  {Koschny}, {Wainscoat}, {Chambers}, {Flewelling}, {Huber}, {Magnier}, \&
  {Weryk}}]{2019Icar..317...39M}
{Micheli}, M., {Cano}, J.~L., {Faggioli}, L., {et~al.} 2019, \icarus, 317, 39

\bibitem[{{Milani} {et~al.}(2000){Milani}, {Chesley}, {Boattini}, \&
  {Valsecchi}}]{2000Icar..145...12M}
{Milani}, A., {Chesley}, S.~R., {Boattini}, A., \& {Valsecchi}, G.~B. 2000,
  \icarus, 145, 12

\bibitem[{{Milani} {et~al.}(2005){Milani}, {Chesley}, {Sansaturio}, {Tommei},
  \& {Valsecchi}}]{2005Icar..173..362M}
{Milani}, A., {Chesley}, S.~R., {Sansaturio}, M.~E., {Tommei}, G., \&
  {Valsecchi}, G.~B. 2005, \icarus, 173, 362

\bibitem[{{Milani} \& {Gronchi}(2010)}]{Milani..book}
{Milani}, A. \& {Gronchi}, G. 2010, Theory of orbits (Cambridge University
  Press)

\bibitem[{{Valsecchi} {et~al.}(2003){Valsecchi}, {Milani}, {Gronchi}, \&
  {Chesley}}]{2003A&A...408.1179V}
{Valsecchi}, G.~B., {Milani}, A., {Gronchi}, G.~F., \& {Chesley}, S.~R. 2003,
  \aap, 408, 1179

\bibitem[{{Veres} {et~al.}(2017){Veres}, {Farnocchia}, {Chesley}, \&
  {Chamberlin}}]{2017..296..139M}
{Veres}, P., {Farnocchia}, D., {Chesley}, S.~R., \& {Chamberlin}, A.~B. 2017,
  \icarus, 296, 139

\end{thebibliography}

\end{document}